\documentclass[12pt, a4paper]{article}

\usepackage{graphicx,amssymb,amsmath,enumerate}

\def\abstract#1{\vskip 7mm
        \begin{center}{\large Abstract}\par \smallskip
                \begin{minipage}[c]{12cm}
                        \small #1
                \end{minipage}
        \end{center}
}
\def\title#1{\begin{center}{\Large\bf #1}\end{center}}
\def\author#1{\vskip 5mm \begin{center}{#1}\end{center}}
\def\address#1{\begin{center}{\it #1}\end{center}}
\makeatletter
\@ifundefined{lesssim}{}{}
\@ifundefined{gtrsim}{}{}
\def\vereq#1#2{\lower3pt\vbox{\baselineskip1.5pt \lineskip1.5pt
\ialign{$\m@th#1\hfill##\hfil$\crcr#2\crcr\sim\crcr}}}
\makeatother

\begin{document}

\title{%
  Asymptotically flat anisotropic space-time in 5 dimensions
  \smallskip \\
  }
\author{%
  Manavendra Mahato\footnote{E-mail: manav@iiti.ac.in},
  Ajay Pratap Singh\footnote{E-mail: phd11115106@iiti.ac.in }
}
\address{%
  $^1,^2$
  Department of Physics, Indian Institute of Technology, Indore\\
  M Block annexe, IET campus, Khandwa Road, Indore -452001, India\\
}
\vspace{4 cm}
\abstract{
  We construct and investigate non conformal anisotropic Bianchi type VII solutions in 5 dimensions. The solutions are asymptotically flat, but they contain a naked singularity at the origin. We also construct solutions of Einstein-Maxwell gravity using the method employed in Majumdar -Papapetrou solutions with various profiles of charged dust. In a fictitious case of negative matter density, we obtain a solution with horizon hiding the singularity.
}
\vspace{10 cm}
\pagebreak
 \def\a{\alpha} \def\as{\asymp} \def\ap{\approx}
\def\b{\beta} \def\bp{\bar{\partial}}
 \def\cA{{\cal{A}}}\def\cD{{\cal{D}}} \def\calO{{\cal{O}}} \def\cL{{\cal{L}}} \def\cP{{\cal{P}}} \def\cR{{\cal{R}}}
 \def\da{\dagger} \def\d{\delta}
\def\D{\Delta}
 \def\e{\eta} \def\ep{\epsilon} \def\eq{\equiv}
 \def\f{\frac}
\def\g{\gamma} \def\G{\Gamma}
 \def\hs{\hspace}
\def\i{\iota}
\def\k{\kappa}
\def\lf{\left} \def\l{\lambda} \def\la{\leftarrow} \def\La{\Leftarrow} \def\Lla{\Longleftarrow}\def\lg{\lgroup} \def\Lra{\Longrightarrow} \def\L{\Lambda}
\def\m{\mu}
\def\n{\nu} \def\na{\nabla} \def\nn{\nonumber}
\def\o{\omega}\def\O{\Omega}\def\ov{\overline}
\def\p{\phi} \def\P{\Phi} \def\pa{\partial} \def\pr{\prime}
\def\r{\rho} \def\ra{\rightarrow} \def\Ra{\Rightarrow}\def\ri{\right} \def\rg{\rgroup}
\def\s{\sigma} \def\sq{\sqrt} \def\S{\Sigma} \def\si{\simeq} \def\st{\star}
\def\t{\theta}\def\ti{\tilde}\def\tb{\tilde{\b}}  \def\tm{\times} \def\tV{\tilde{V}} \def\tr{\textrm} \def\T{\Theta}
 \def\u{\upsilon}
\def\U{\Upsilon}
\def\v{\varepsilon} \def\vh{\varpi} \def\vk{\vec{k}} \def\vp{\varphi} \def\vr{\varrho} \def\vs{\varsigma}\def\vt{\vartheta}
 \def\w{\wedge}
 \def\z{\zeta}
\newcommand{\be}{\begin{equation}} \newcommand{\ee}{\end{equation}}
\newcommand{\bea}{\begin{eqnarray}} \newcommand{\eea}{\end{eqnarray}} 
\newcommand{\bet}{\begin{tabular}}\newcommand{\eet}{\end{tabular}}
\newcommand{\bay}{\begin{array}}\newcommand{\eay}{\end{array}}
\section{Introduction}
Understanding of nature using physics has revealed its richness and beauty in many ways. General relativity encodes the interaction of matter or radiation with space time in a charming manner. Though it is observed that many anisotropic solutions do exist in nature, spherically symmetric solutions have received more focus till now because of their simplicity.
 Homogenous, but anisotropic solutions of Einstein equations were classified by Bianchi many decades ago.\cite{LLif, RS} However, finding such solutions have always been challenging and has led to some numerical results previously. Recently, anisotropic space times have gathered a renewed interest from a different direction. Many solutions of general relativity are much sought for the study of field theory using AdS/CFT correspondence as well as attractor mechanism.\cite{Iizuka:2012iv, Iizuka:2012pn, Kachru:2013voa}, Recently , asymptotic AdS (anti-de Sitter), anisotropic solutions were constructed and studied to investigate properties of certain anisotropic condensed matter systems using AdS/CFT correspondence. Anisotropic spacetimes are interesting systems and may contain many unique properties. Given the difficulty in their construction in 4 dimensions, and recent constructions of anisotropic solutions in 5 dimensions hint that that it may be less restrictive to construct anisotropic solutions in higher dimensions.\\

  Asymptotically AdS, Bianchi $VII$ class of solutions has received much attention as they were related to spatially modulated superconductors where the Cooper pair is not an s-wave, but a p-wave. The numerical solutions were constructed and their properties were further studied.\cite{Nakamura:2009tf, Ooguri:2010kt, Donos:2012gg, Donos:2012wi} This motivated us that similar anisotropic solutions with asymptotically flat or de Sitter property can also be constructed. We take up the investigation of such solutions in this manuscript. \\

  In this manuscript, we report certain analytic Bianchi $VII_0$ solutions, but they contain a naked singularity. We further search for anisotropic solutions of Einstein Maxwell theory and construct few such solutions which can be considered as generalizations of Majumdar Papapetrou solutions for our case.\cite{Myers:1986rx, Gibbons:2008hb, Varela:2002vp, Frolov:2012jj} We seek to construct solutions with a regular horizon hiding any naked singularity and we report one such case, though it is sourced by a fictitious matter density.

\section{Bianchi Classes}
Bianchi classified three dimensional homogeneous spaces into different classes. We present its brief review in this section. As we know that homogeneity means identical metric properties at all points of the space. Mathematically, homogeneity means the form of the metric does not change under translations. Translations along all directions are isometries and they are generated by the Killing vectors written in notations of differential geometry as
\begin{flalign}
X_{a}&=e^\a_{a}\f{\pa}{\pa x^\a}
\end{flalign}
If we do not assume isotropy, then these Killing vectors will not commute in general.\\
The commutators of generators can be represented as
\begin{flalign}
[X_{a},X_{b}]&=C^c_{ab}X_{c}
\end{flalign}
where $C^c_{ab}$ are structure constants. They are antisymmetric in lower indices i.e. \\
\be C^c_{ab}=-C^c_{ba}\ee
and they also satisfy Jacobi identity,
\be C^e_{ab}C^d_{ec}+C^e_{bc}C^d_{ea}+C^e_{ca}C^d_{eb}=0.\ee
The structure constants are written in a form which separates its symmetric and antisymmetric parts as follows,
\be
C^c_{ab}=\ep_{abd}n^{dc}+\d^c_{b}a_{a}-\d^c_{a}a_{b}
\ee
where,
$\ep_{abd}$ is the unit antisymmetric tensor,
$n^{ab}$ is a symmetric tensor with eigenvalues $n^{(1)}$, $n^{(2)}$ and  $n^{(3)}$ and
$a_{a}$ is a vector. Next, a choice of frames is made with eigenvalues of $n_{ab}$ chosen as the basis. The vector $a_{a}$ is then chosen to be along a certain direction i.e. $\vec{a}=(a,0,0)$.
The Jacobi identity gets reduced to
$n^1a=0$. Thus, either $a$ or $n_1$ has to vanish.
The commutators of generators can then be explicitly written as
\begin{flalign}
[X_{1},X_{2}]&=aX_{2}+n^{(3)} X_{3},\nn\\
[X_{2},X_{3}]&=n^{(1)} X_{1},\nn\\
[X_{3},X_{1}]&=n^{(2)} X_{2}-aX_{3}.
\end{flalign}
Using scale transformations and choice of signs, the list of possible types of homogeneous spaces turn out to be
\begin{center}
\begin{tabular}{|l|l|l|l|l|}
\hline
type      & a   & n1    & n2      & n3\\
\hline
I       & 0     & 0    & 0      & 0\\
II      & 0     & 1    & 0      & 0\\
III     & a     & 0    & 1      & -1\\
IV      & 1     &0     &0       &1\\
V       & 1     &0&0&0\\
VI      & a     &0&1&-1\\
VII     & a     &0&1&1\\
VIII    & 0     &1&1&-1\\
IX      & 0     &1&1&1\\
\hline
\end{tabular}
\end{center}
We next choose to work with Bianchi type $VII_0$, which chooses $a$ and $n^i$ both to be zero. We explicitly write the three vectors below.
\bea
X_1&=&\pa_1,\nn\\
X_2&=&\cos(kx_1)\pa_2+\sin(kx_1)\pa_3,\nn\\
X_3&=&-\sin(kx_1)\pa_2+\cos (kx_1)\pa_3.
\eea
 Henceforth, we will investigate solutions containing such anisotropy.
\section{Anisotropic solution of pure gravity action}
We consider a pure gravity action in five dimensions. It is known to be hard to construct solutions in 4 dimensions. But, recent progres have shown that the restrictions are little relaxed in higher dimensions. Our action is
\be
S=\int \sqrt{|g|}(R)
\ee
Here, $R$ is the Ricci scalar and $g$ denotes determinant of the metric. The signature of our metric is (-1,1,1,1,1). We choose our ansatz for the metric to be
\be
ds^2= -e^{2T(r)}dt^2+dr^2+e^{2M(r)+2N(r)}\o_1^2+e^{2M(r)-2N(r)}\o_2^2+e^{2Z(r)}\o_3^2,
\ee
where, $T(r)$, $M(r)$, $N(r)$ and $Z(r)$ are chosen to be functions of $r$ only. The one forms used above are
\bea
\o_1&=&\cos(kx_1)dx_2+\sin(kx_1)dx_3,\nn\\
\o_2&=&-\sin(kx_1)dx_2+\cos (kx_1)dx_3,\nn\\
\o_3&=&dx_1.
\eea
Thus, we are looking for static, anisotropic Bianchi $VII_0$ solutions since we are restricting the appearance of cordinates $x_2$ and $x_3$ in the metric in the above specfic combinations only. We choose to work in a non-coordinate basis with the following vielbeins. The metric in such a basis is diagonal. Our vielbeins are

\begin{flalign*}
e^t &=e^{T(r)}dt,\\
e^r &=dr,\\
e^a &=e^{M(r)+N(r)}\o_1,\\
e^b &=e^{M(r)-N(r)}\o_2,\\
e^c &=e^{Z(r)}\o_3.
\end{flalign*}

We get the following Einstein equations of motion 
\begin{flalign}
T''G'&=0,\nn\\
M''G'&=0,\nn\\
N''G'-k^2e^{-2Z}\sinh{4N}&=0,\nn\\
Z''G'+2k^2e^{-2Z}\sinh^2{(2N)}&=0,\nn\\
G''^2+2k^2e^{-2Z}\sinh^2{(2N)}&=0,
\end{flalign}
including a constraint
\begin{flalign}
{G'}^2+({T'}^2+2{M'}^2+2{N'}^2+{Z'}^2 )+\f{k^2}{2}(e^{(2N-Z)}-e^{(-2N-Z)})^2=0.
\end{flalign}
Here, variable $G$ denotes $T+2M+Z$ and superscript prime denotes derivative with respect to $r$. The first two equations suggest that a good radial variable will be $u$ defined as
\begin{flalign}
du & =e^{-G}dr.
\end{flalign}
Also, we define a different variable $L=G-Z$. Then, the set of equations are
\begin{flalign}
T_{uu}&=0,\nn\\
M_{uu}&=0,\nn\\
L_{uu}&=0,\nn\\
N_{uu}-k^2e^{2L}\sinh{4N}&=0,\nn\\
Z_{uu}+2k^2e^{2L}\sinh^2{2N}&=0,\nn\\
G_{uu}+2k^2e^{2L}\sinh^2{2N}&=0.
\end{flalign}
The constraint equation now becomes
\be
G_u^2+T_u^2+2M_u^2+2N_u^2+Z_u^2+2k^2e^{2L}\sinh^2(2N)=0.
\ee
We then proceed to express the functions $T$, $M$ and $L$ as
\bea
T&=t_0+2t_1u,\nn\\
M&=m_0+m_1u,\nn\\
L&=a+l_1u,
\eea
where $t_0$, $t_1$, $m_0$, $m_1$, $a$ and $l_1$ are constants.
We find the next amenable equation to solve is that for $N$. However, this equation become simpler if one chooses the parameter $l_1$ to be $0$ or function $L(r)$ to be a constant function.
The equation for $N$ then reduces to
\be
N_{uu}-\l^2\sinh(4N)=0,
\ee
where $\l=ke^a$ is a constant. Inverting this differential equation results in
\be
\f{u_{NN}}{u_{N}^3}+\l ^2\sinh(4N)=0,
\ee
which if integrated once, leads to
\be
\l^2u_N^2=\f{2}{(c_1+\cosh(4N))}.
\ee
Here, $c_1$ is an integration constant. The above equation can be solved in terms of Jacobi amplitudes. However, it offers simple solution for two cases, (1)$c_1=-1$ and (2)$c_1=1$. We next proceed to explore the case of  $c_1=-1$ in more detail. We can then write the above equation as
\be\label{Nueqn}
\l u_N=\f{1}{\sinh(2N)}{\textrm{\hspace{1 cm}or\hspace{1 cm}}}N_u=\l \sinh(2N).
\ee
Its general solution is
\be
e^{-2N}=\tanh(\l u+u_0).
\ee
It can also be written as
\be
2\sinh [2(\l u+u_0)]\sinh (2N)+1=0.
\ee
The equation for $Z$ can be written as
 \be
 Z_{uu}=-\f{2\l ^2}{\sinh^2 [2(\l u+u_0)]}.
 \ee
 It has the general solution
 \be
 Z=Z_0+Z_1 u+\f{1}{2}\ln\sinh [2(\l u+u_0)].
 \ee
Since $L$ was taken as a constant and $G=L+Z$, variable $G$ also gets determined.
Most of the free parameters get fixed by the constraint relation, which now takes the form
\be
G_u^2-(T_u^2+2M_u^2+2N_u^2+Z_u^2)+\f{k^2}{2}e^{2a}\sinh^2(2N)=0.
\ee
Since, $G_u=Z_u=T_u+2M_u+Z_u$, we get $t_1=-2m_1$. One can deduce using eqn. (\ref{Nueqn}) that $m_1^2=0$. Further, we can absorb the parameters $Z_0$,  $t_0$ and $Z_1$ by redefining coordinates $x_3$, $t$ and $u$ respectively. Furthermore, variable $m_0$ can be absorbed by redefining coordinates $x_2$ and $x_3$. Moreover, redefining the constant a, the metric can be written in a form
\bea\label{5dmetric}
ds^2=-a^2dt^2+e^{2u}\sinh{(2\l u)}(a^2du^2+\o_3^2)+\f{1}{\tanh{(\l u)}}\o_1^2+\tanh{(\l u)}\o_2^2.\nn\\
\eea
This metric is Ricci flat  ($R_{\mu\nu}=0$) but contains a naked singularity at the origin. The Kretschmann tensor blows at $u=0$. However, we notice that the metric component along the time direction is just a constant. In fact, we have a spatial 4 dimensional submanifold with a naked singularity in addition with a time direction. We hope that the naked singularity can still be put in a physical context if we excite some other field whose energy density itself becomes infinity at $u=0$ thus causing strong curvature there. If it happens, there is hope that this singularity can be hidden behind a horizon.

\section{Anisotropic solution of Einstein Maxwell action}
We will now try to construct anisotropic solutions of 5 dimensional Einstein Maxwell action along with matter density. Our action in this section is
\be
S=\int d^5x \sqrt{-g}\lf [R-\f{1}{4}F_{\m\n}F^{\m\n}+A_{\m}J^{\m}-\f{\r}{2}(g^{\m\n}u_{\m}u_{\n}+1)\ri ].
\ee
Here, notation $g$ denotes the determinant of the metric which we choose to have one negative signature along time direction as earlier. Ricci scalar is denoted by $R$. The electromagnetic potential and field strength are denoted by $A_{\m}$ and $F_{\m\n}$, respectively. We also incorporate source for electromagnetic field denoted as $J_{\m}$ as well as a matter density denoted by $\r$ with a velocity $u_{\m}$. We take the velocity to be non dynamical field. The matter term in action is conspired to give the correct energy momemtum tensor for a pressureless dust i.e. $T_{\m\n}=\r u_{\m}u_{\n}$. The Maxwell equation is
\be\label{Maxwell1}
\na_{\n}F^{\m\n}=J^{\m}.
\ee
The metric fluctuations of the action leads to following Einstein equations.
\be
R_{\m\n}=\f{1}{2}F_{\m\r}{F_{\n}}^{\r}-\f{1}{12}g_{\m\n}F_{\r\s}F^{\r\s}+\r u_{\m}u_{\n}+\f{1}{3}g_{\m\n}\r .
\ee
We further attempt to find solutions of these set of equations using a method employed earlier to find generalizations of Majumdar-Papapetrou metrics.\cite{Varela:2002vp, Frolov:2012jj, Gurses:1998zu}
We mention that Majumdar Papapetrou metrics are 4 dimensional extremal solutions of Einstein-Maxwell equations of the type
\be
ds^2=-V(\vec{x})^2dt^2+\f{1}{V(\vec{x})^2}(d{x_1}^2+d{x_2}^2+d{x_3}^2)
\ee
alongwith an electromagnetic flux of the kind $F =d(V^{-1})$. The function V is required by Einstein equations to be a harmonic function of the 3 dimensional flat subspace. When searching for solutions in (m+1) dimensions, we generalize the metric ansatz to be of type
\be
ds^2=-V(x_i)^2 dt^2+\f{1}{V^{2/n}(x_i)} h^{ij} dx_i dx_j \hspace{2 cm}{i=1,2,...,m.}
\ee
The metric $h_{ij}$ depends on spatial coordinates only. It leads to the following Ricci tensor,
\bea
R_{tt}&=&V^{(1+2/n)}\na ^2_{(h)}V-\f{m-2}{n}V^{2/n}(\na V)^2\nn\\
R_{ij}&=& R^{(h)}_{ij}-\f{(mn+2-m)}{n^2V^2}\pa _iV\pa _jV+h_{ij}\lf [\f{\na ^2_{(h)}}{nV}-\f{(m-2)}{n^2}\lf(\f{\na V}{V}\ri )^2\ri]\nn\\
&&+\f{(m-2-n)}{nV}\na_i\na_jV.
\eea
The indices ${i,j}$ denote spatial coordinates only.  Here, notation $\na^2_{(h)}$ denotes Laplacian defined over the internal space with metric $h_{ij}$. The Ricci tensor component $R_{ti}$ is found to be vanishing.
When we write the corresponding  energy momentum tensor $T_{\m\n}$ and try to satisfy Einstein equations, we notice that there is no analog of any term like $\na_i\na _jV$ in the expression of $T_{\m\n}$. Such a term, if kept, will require us to solve complicated non-linear equations. One chooses a relation between parameters $m$ and $n$, so as to make such term vanish i.e. $n=m-2$. Returning to our interest of 5 dimensional metrics, we find that we should take $m=4$ and $n=2$. Thus, our metric ansatz reduces to
\be
ds^2=-V(x_i)^2 dt^2+\f{1}{V(x_i)} h^{ij} dx_i dx_j
\ee
We next evaluate the components of the Einstein tensor and they are found to be
\bea
G_{00}&=&\f{3}{2} V^2 \nabla _h^2 V-\f{9}{4} V h^{ij} \partial_i V \partial_j V+\f{1}{2} V^3 R_{(h)}\nn\\
G_{ij}&=&R_{(h)ij}-\f{3}{2 V^2}  \partial_i V \partial_j V+\f{3}{4 V^2}  h_{ij} h^{kl} \partial_k V \partial_l V-\f{1}{2} h_{ij} R_{(h)}.
\eea
We choose our internal subspace to be the anisotropic Ricci flat space that we obtained in last section. Thus, Ricci tensor $R_{(h)}$ and Ricci scalar $ R_{(h)ij}$ vanishes. The Einstein tensor in our case then reduces to
\bea
G_{00}&=&\f{3}{2} V^2 \nabla _h^2 V-\f{9}{4} V h^{ij} \partial_i V \partial_j V\nn\\
G_{ij}&=&-\f{3}{2 V^2}  \partial_i V \partial_j V+\f{3}{4 V^2}  h_{ij} h^{kl} \partial_k V \partial_l V
\eea
Next we make an ansatz for the electromagnetic potential $A_{\m}$. We assume it to be along the time direction
\be
A_{\m}=A\d^{0}_{\m}.
\ee
We also make an ansatz for the four velocity of the matter density. We assume matter to be at rest i.e.
\bea
u_{\m}&=V\delta ^{0}_{\m}
\eea
Such a choice also ensures that $u_{\m}u^{\m}=-1$.
The energy momentum tensor component from electromagnetic field is given in terms of field strength tensor $F_{\m\n}=\pa _{\m}A_{\n}-\pa _{\n}A_{\m}$ as
\bea
T^{field}_{\m\n}&= \f{1}{4 \pi}(F_{\m\r} {F_{\n}}^{\r} -\f{1}{4} g_{\m\n} F_{\r\s} F^{\r\s}).
\eea
The matter energy density contribution to energy momentum tensor is
\bea
T^{matter}_{\m\n}&=\rho u_{\m} u_{\n}.
\eea
Thus the components of total energy-momentum tensor turn out to be
\bea
T_{00}&=&\f{1}{3} V h^{ij} \partial_i V \partial_j V+\f{2}{3}\rho V^2\nn\\
T_{ij}&=&- \f{1}{2V^2} \pa_i V \pa_j V+\f{h_{ij}}{6 V^2}  \{(\na _h A)^2+2\r V\}
\eea
The non trivial components of Einstein equations are
\bea
\label{eqn1}
V^2 \nabla _h^2 V-V (\na _hV)^2&=&\f{1}{3} V (\na _h A)^2+\f{2}{3} \rho V^2\\
\label{eqn2}
\f{3  \partial_i V \partial_j V}{2 V^2}-\f{h_{ij}}{2V^2}\lf \{ V\na ^2_h V-(\na V)^2\ri \}&=&
\f{\partial_i A \partial_j A}{2V^2}-\f{h_{ij}}{6V^2}\lf \{(\na _h A )^2+2\r V\ri \}
\eea
We notice that the term explicitly proportional to $h_{ij}$ in equation (\ref{eqn2}) is same as $(tt)$ component Einstein  equation as in (\ref{eqn1}).  Canceling it, we get
\be
\f{3  \partial_i V \partial_j V}{2 V^2}=\f{  \partial_i A \partial_j A}{2V^2}.
\ee
It can be solved easily if we take $A$ proportional to $V$.  The equation fixes the relation to be
$A=\sqrt{3}V$. Then the rest of Einstein equations simplifies to
\be
\na _h^2\lf (\f{1}{V}\ri )=-\f{2\r}{3V^2}.
\ee
In terms of $\l =\f{1}{V}$, the above equation can be written as
\be\label{Le}
\na _h^2\l =-\f{2\r\l ^2}{3}.
\ee
We next make an ansatz for the source of the electromagnetic field. We take only the time component of $J_{\m}$ to be non trivial. Alongwith the above choice for electromagnetic potential, the Maxwell equation takes a form
\be
\na_h^2\l=-\f{\l J^t}{\sqrt{3}}.
\ee
This equation will be consistent with the equation (\ref{Le}), if we chose
\be
J^t=\f{2\r\l}{\sqrt{3}}
\ee
Thus we are left with a single equation viz. equation (\ref{Le}), which is a non homogenous Laplacian equation. We next proceed to solve it for some suitable choices of matter density.

\section{Anisotropic solutions with chosen sources}
\subsection{Polynomial solutions}
The equation we need to solve is
\be\label{Feqn}
\na _h^2\l =\f{1}{\sqrt{h}} \partial_i (\sqrt{h} h^{ij} \partial_j \l )=-\f{2\r\l ^2}{3}.
\ee
We choose our spatial subspace to be same as the anisotropic 4 dimensional subspace which was obtained in the second section. Therefore,
\be\label{4dmetric}
h_{ij}dx^idx^j=e^{2u}\sinh{(2\l u)}(a^2du^2+dx_1^2)+\f{1}{\tanh{(\l u)}}\o_1^2+\tanh{(\l u)}\o_2^2.
\ee
The determinant of the metric is $a e^{2u}\sinh{(2\l u)}$. We will henceforth denote $x_1$ simply by $x$. For simplicity, we assume the function $\l$ to be a function of $u$ and $x$ only.

Then equation (\ref{Feqn}) becomes,
\bea
\f{1}{\sqrt{h}}\lf (\f{\partial_{u}^{2} \l}{a}+a\partial_{x}^{2} \l\ri )+\f{2}{3}\rho \l^2 &=0
\eea
We next define polar coordinates $au=r\cos\p $ and $x=r\sin\p$.
The equation appears in polar form as
\bea\label{finalE}
\partial_{r}^{2} \l+\f{1}{r}\partial_{r} \l &=-\f{2}{3a} \sqrt{h} \rho \l^2
\eea
where we have taken $\l$ to be independent of $\phi$, i.e. we restrict ourselves to the lowes harmonic.
Now the equation can be made amenable to analytical results for suitably choosing the profiles for the matter density. We next choose
\bea
\rho &=\f{3a c}{2 r^n \sqrt{h}  \l^2}
\eea
where, $n$ is a positive integer greater than $2$ and $c$ is a constant. Such a form of energy density is physically reasonable as it vanishes smoothly to zero when one proceeds towards infinity. Then equation (\ref{finalE}) becomes,
\bea
\partial_{r}^{2} \l+\f{1}{r}\partial_{r} \l &=-\f{c}{r^n}
\eea
One can easily solve it to obtain
\bea
\l &=-\f{c}{(n-2)^2r^{n-2}}+c_2+c_1 \log{r}.
\eea
We can restrict ourselves to polynomial form by choosing $c_1=0$. This leads to
\bea
\l &=c_2-\f{c}{(n-2)^2r^{n-2}}.
\eea
By choosing $n=3$, we get
\bea
V &=\f{1}{c_2-\f{c}{r}}
\eea
Then, the metric now appears as
\bea
ds^2&=-\f{1}{(c_2-\f{c}{r})^2} dt^2+(c_2-\f{c}{r}) h^{ij} dx_i dx_j.
\eea
But, this solution shows two essential singularity where Kretschmann tensor diverges. They are $r=0$ and $r=c/{c_2}$. Thus there are two naked singularities in this solution. We next consider a fictitious matter whose density profile is negative by replacing the constant $c$ with $-C$. This sends the second singularity at $r=c/(c_2)$ to a negative value of $r$, thus out of the spacetime. Choosing constant $c_2=1$,  the metric now appears as
\be
ds^2=-\f{1}{(1+\f{C}{r})^2} dt^2+(1+\f{C}{r}) h^{ij} dx_i dx_j.
\ee
We find that the $(tt)$ component of the metric for small values of $r$ is
\be
g_{tt}=\f{1}{\lf(1+\f{C}{r}\ri )^2}\sim 1-\f{2C}{r}.
\ee
 Thus, this solution has a horizon near $r\sim 2C$, which also hides the essential singularity residing at $u=0$. We expect this solution to be of extremal type as the same is true for all such previous generalizations of Majumdar Papapetrou metrics.
\subsection{The Sine-Gordon Solution}
One can get a Sine Gordan kind of equation here by a different choice of matter density. First we choose a different radial coordinate
\bea
\tau = \log{r} \nn
\eea
Then the equation (\ref{finalE}) becomes,
\bea
\partial_{\tau}^{2} \l &=-\f{2}{3} e^{2\tau} \sqrt{h} \rho \l^2
\eea
Next, we choose matter density profile to be of form
\bea
\rho &=\f{3 a\delta^2 \sin{\l}}{2 e^{2\tau} \sqrt{h}  \l^2}
\eea
The above equation then reduces to a Sine-Gordan equation,
\bea
\partial_{\tau}^{2} \l +\delta^2 \sin{\l} &=0.
\eea
It admits a solution
\bea
\l(\tau)&=4 \arctan{\lf [\tanh{\lf (\f{\delta \tau +c}{2}\ri )}\ri ]}
\eea
This ensures that $g_{tt}$ is finite everywhere except at origin $r=0$, where one encounters a naked singularity.
\section{Conclusion}
We have constructed an anisotropic 4 dimensional asymptotically flat Riemannian metric which
 is Ricci flat. We later incorporated it in a 5 dimensional space time along with matter density and electro magnetic flux using a method similar to Majumdar Papapetrou way of constructing extremal solutions. With certain choices of  matter density profiles, we were able to construct explicit solutions. Most of them contain the naked singularity. However, with a fictiticious choice of matter density, we arrived at a solution with a horizon hiding the naked singularity. Given a Laplacian equation with a negative source, we are more likely to come across naked singularity. For a horizon to appear, we generally need $g_{tt}$ to vanish at some finite radial coordinate. This requires function $V$ to vanish and its inverse, $\l$ to blow up at some $r$. However, by a suitable choice of variables, we can write the Laplacian as a second derivative of $\l$. For a second derivative to be negative at all positions as dictated by its negative source, the function $\l$ becomes a convex function. A function blowing up at a certain finite value generally should  be concave instead of convex. Thus, when we take the source to be positive by taking the matter density negative, we arrive at a spacetime with horizon. Using our method, interesting anisotropic solutions in higher dimensions may be constructed by further investigating further types of Lagrangians which can hide the naked singularity present in the center of the subspace behind a horizon.

\end{document}